\documentstyle[times]{jaa}
\include{graphicx}

\begin{document}

\title[Radio Band Variability]
      {Radio Band Observations of Blazar Variability} 
\author[Aller, Aller, \& Hughes] 
       {Margo F.~Aller, Hugh D.~Aller, \& Philip A.~Hughes \\ 
        University of Michigan, Ann Arbor, MI, 48109-1042, USA}
\maketitle

\label{firstpage}
\begin{abstract}
The properties of blazar variability in the radio band are studied
using the unique combination of temporal resolution from
single dish monitoring and spatial resolution from VLBA imaging; such measurements,
now available in all four Stokes parameters, together with
theoretical simulations, identify the origin of radio band
variability and probe the characteristics of the radio jet where
the broadband blazar emission originates. Outbursts in total flux
density and linear polarization in the optical-to-radio bands
are attributed to shocks propagating within the jet spine, in part
based on limited modeling invoking transverse shocks; new 
radiative transfer simulations allowing for
shocks at arbitrary angle to the flow direction confirm this
picture by reproducing the observed centimeter-band
variations observed more generally, and are of current interest
since these shocks may play a role in  the $\gamma$-ray flaring
detected by {\it Fermi}. Recent UMRAO multifrequency Stokes V studies of 
bright blazars identify the spectral variability properties
of circular polarization for the first time and demonstrate
that polarity flips are relatively common.  All-Stokes data
are consistent with the production of circular polarization by
linear-to-circular mode conversion in a region that is at least
partially self-absorbed. Detailed analysis of single-epoch, multifrequency,
all-Stokes VLBA observations of 3C~279 support this physical picture
and are best explained by emission from an electron-proton plasma.Ê
\end{abstract}


\section{Overview}
\label{sec:over}
In this review we discuss the properties of centimeter-to-millimeter band
variability in Stokes I (total flux density), and compare the derived
values to those determined in the {\it Fermi} $\gamma$-ray band. 
We summarize evidence for the shock-in-jet model invoked for explaining the optical-to-radio-band 
variations, and present new modeling results allowing for oblique shocks. 
We show Stokes V (circular polarization) light curves illustrating the
range of behavior found, and illustrate how  such measurements can be used
in combination with all-Stokes, multifrequency imaging to place constraints on jet properties.

\begin{figure}
 \includegraphics[width=10cm,height=6.0cm, clip=true]{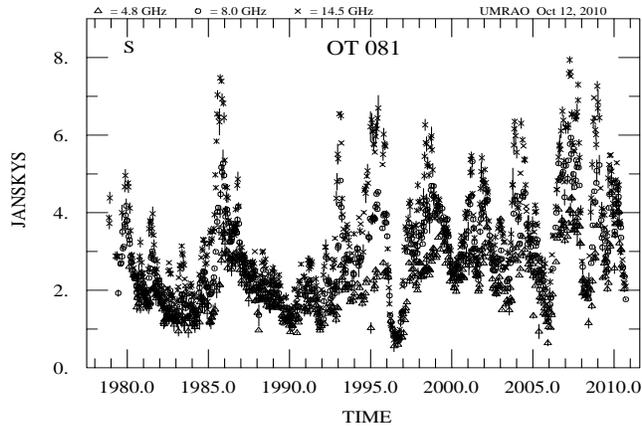}
\caption{Two-week averages of total flux density observations over 3 decades illustrating
the characteristic behavior of radio band variability. Complementary 15 GHz MOJAVE VLBA imaging observations
indicate a very compact source (e.g. Kovalev et al. 2005).}
\end{figure}

\section{Total Flux Density: Stokes I} 

Single dish monitoring observations from Mets\"ahovi, Michigan (hereafter UMRAO), and
recently from the OVRO 15 GHz program, probe the properties of variability in the centimeter-to-millimeter
bands on timescales of days to decades. In very active sources,
e.g. see Figure 1, the variability in Stokes I is continuous (no quiescent periods).
Events do  not repeat in either amplitude or spacing, and the amplitude
changes by at most a factor of $\sim$8.
In the spectral range 14.5 to 4.8 GHz observed by UMRAO,
variations are highest at 14.5 GHz, and lowest and often delayed at 4.8 GHz.
While the outbursts do not exhibit true periodicities, quasi-periodicities
on the order of a few years been identified in many sources; 
plausibly these correspond to the dynamical response time
of the flow to perturbations. Doppler factors have been determined
from fits to the fastest events, and values generally lie within the
range $0\leq$D$_{var}\leq35$ (e.g. Hovatta et al. 2009);
however, as noted by Fan et al. (2009), there is considerable spread
in the results obtained for a specific source when different analyses
are compared. Structure functions have been used to identify both a characteristic
time scales and the noise process responsible for the variability; typically
the time scales are two years with some spread, and the noise process
is shot noise (Hughes, Aller, \& Aller 1992). These radio band properties are
characteristically different from those identified applying similar analysis procedures
to {\it Fermi} monitoring data; there the emission process
is almost always flicker, the variability time scales are typically 
4 to 12 weeks, and no periodicity is identified. (Adbo et al. 2010)

\begin{figure}
 \includegraphics[width=8cm,height=10cm, clip=true]{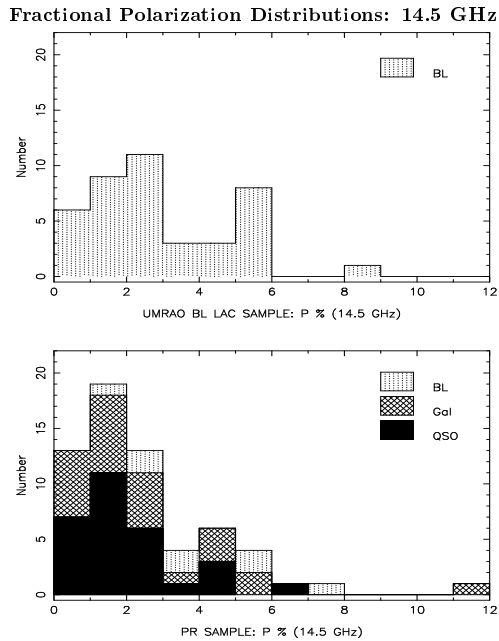}
\caption{Histograms of fractional polarization determined from time-averaged long-term Q and U measurements
for two flux-limited samples: the UMRAO BL Lac sample (1979-2005) and the Pearson-Readhead sample (1984-2005).}
\end{figure}

\section{Origin of the Variability}
\label{sec:origin}

\begin{figure}
\includegraphics[width=9cm,height=7.5cm, clip=true]{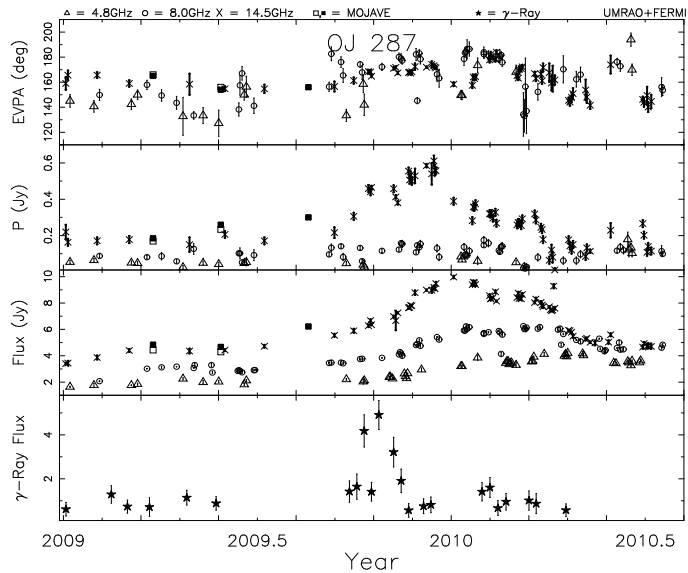}
\caption{Example of a source exhibiting an oblique shock signature during a $\gamma$-ray flare.
Linear polarization and total
flux light curves are shown in panels 2-4. The lower panel gives the $\gamma$-ray light curve
kindly provided by S. Jorstad (units: photons/sec/cm$^2$x10$^{-7}$).}
\end{figure}
\begin{figure}
\includegraphics[width=8cm,height=5cm, clip=true]{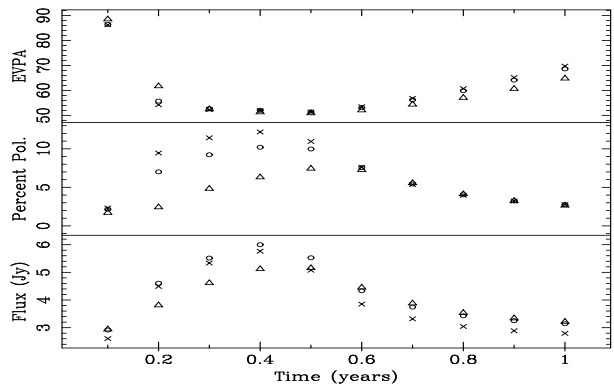}
\caption{Simulated light curves at 3 frequencies assuming
a forward moving shock oriented at an obliquity of 45$^{\circ}$, and a shock compression of 0.7.}
\end{figure}

The generally-accepted scenario for the production of the radio band variability is that shocks develop
naturally within the relativistic jet flows (e.g. Hughes 2005). The magnetic field within the emitting
region is turbulent. Supporting evidence comes from the low
degrees of time-averaged linear polarization (see Figure 2), which are
much less than 60-70\% expected for a canonical synchrotron source. The shock
produces a compression as it passes through the emitting region
yielding an increased degree of order.  The event signature is a
monotonic swing in the electric vector position angle (EVPA) and an increase in the fractional
linear polarization. In our earlier work, the propagating shocks were assumed 
to compress transversely to the flow direction, and fits to events in
three carefully-selected sources successfully reproduced the observed spectral evolution
in both total flux density and linear polarization (e.g. Hughes, Aller, \& Aller 1991).
However, later events in these same sources could {\it NOT} be fit with the same source parameters,
and the observed behavior suggested that more generally shocks are oriented at oblique angles to the flow direction.

As part of an investigation of the role that shocks may play in the generation of $\gamma$-ray flaring observed by {\it Fermi},
we are intensively monitoring a core group of about two dozen $\gamma$-ray and radio bright blazars to determine
whether the shock signature is present during flares detected by {\it Fermi}. We have identified several cases where
this signature is present, and an example is shown in Figure 3. Note that the EVPA exhibits a swing through about
40$^{\circ}$ rather than  90$^{\circ}$ as expected for a transverse shock. As part of this investigation,
new radiative transfer models have been developed which allow for the propagation of shocks at arbitrary
angle to the flow direction. The new models build on
our work carried out in the mid 1980s and are determined primarily by two free
parameters: the shock compression and the shock direction (forward or reverse).
In the example shown in Figure 4 the viewing angle =10$^{\circ}$, the Lorentz factor of the flow=2.5,
and the Lorentz factor of the shock=6.7.
Comparison with Figure 3 shows that the simulation successfully reproduces
the main features of the data: a total flux outburst, an increase in linear polarization to near 10\%, a swing in
EVPA through about 40$^{\circ}$, and the spectral behavior during outburst evolution.

\section{Stokes V: Circular Polarization}
\label{sec:StokesV}
Surveys have shown that Stokes V emission from AGN is common, but this emission has not been observed routinely 
because of the difficulty in detecting this weak emission. VLBA observations place
the CP emission site at or near the radio core (e.g. Homan \& Wardle 2004), the region believed
to be the $\tau$=1 surface or a standing shock. UMRAO monitoring observations of Stokes V were carried out
for a few sources at 4.8 and 8 GHz during the late 1970s and early 1980s, but the program was
subsequently dropped because it restricted us to measurements of a few very bright AGN.
However, in 2002 we resumed the program, adding observations at 14.5 GHz in late 2003.
Specific observational goals are: 1) to ascertain whether the polarity is stable
to test the viability of models such as Ensslin's (2003) whereby the handedness (polarity) is an indicator of
the direction of rotation of the quasar engine; 2) to determine variability time scales; and
3) to look for relations between the variations in all four Stokes parameters as tests of
competing proposed emission mechanisms. We show example data in Figure 5. 

\begin{figure}
\includegraphics[width=11cm,height=8cm, clip=true]{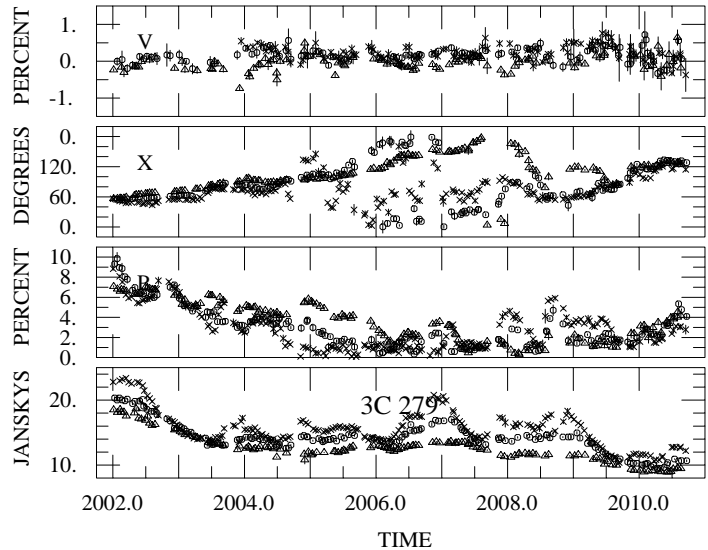}
\caption{Two-week averages of total flux density, LP, and CP since 2002 in 3C~279. This source has exhibited
 repeated polarity flips at 4.8 GHz, and polarity differences as a function of frequency at a single epoch, e.g. in 2004.0.}
\end{figure}

While we find preferred polarities in each source, there are
also polarity flips at 4.8 and 8 GHz in many sources; these
persist for a few months to a few years and commonly occur at
times at which there is evidence for self-absorption in the Stokes I spectrum.
Identification of variability on these long timescales requires a dedicated long-term program rather
than ad hoc measurements. The data are overall consistent with linear-to-circular
mode conversion.

As a next step in our work, we are collaborating with D.~Homan, J.~Wardle, \& M.~ Lister
to obtain multifrequency, all-Stokes VLBA imaging data for a small group of sources.
Stokes V spectra were presented for 3C~84, 3C~279, and 3C~380 in Homan et al.
(2006), and a detailed analysis of the spectra of the three core components responsible for
the CP emission in 3C~279 has been published (Homan et al. 2009).  A self-consistent model
was proposed which provided an estimate of the low energy cutoff of the particle
energy distribution, suggested that a turbulent magnetic
field and not a large scale helical field, plays the dominant role in the mode conversion process, and was
consistent with the presence of a predominantly electron-proton plasma. Data from two additional observing
epochs are not yet analyzed, and it will be very interesting to see whether the same generic picture
is provided by these data.

This work was made possible by support from NSF grant NSF0607523, NASA Fermi grants NNX09AU16G and NNX10AP16G, and by the
U. of Michigan.

\end{document}